\documentclass[aps,prl,twocolumn,superscriptaddress,citeautoscript]{revtex4-1}
\usepackage{amsmath}
\usepackage{color}
\usepackage{psfrag}
\usepackage{amsfonts}
\usepackage{amssymb}
\usepackage{amsbsy}
\usepackage{latexsym}
\usepackage{epsfig}
\usepackage{graphicx}
\usepackage{dcolumn}
\usepackage{bm}
\usepackage{subfigure}
\usepackage{mathtools}
\usepackage[colorlinks,urlcolor=blue,citecolor=blue]{hyperref}
\usepackage{lipsum}

\begin{document}
\title{Quench Dynamics in a Trapped Bose-Einstein Condensate with Spin-Orbit Coupling}
\author{Sheng Liu}
\affiliation{Key Laboratory of Quantum Information, University of Science and Technology of China, Hefei, 230026, China}
\affiliation{CAS Center for Excellence in Quantum Information and Quantum Physics, University of Science and Technology of China, Hefei, 230026, China}
\author{Yongsheng Zhang}\email{Email: yshzhang@ustc.edu.cn}
\affiliation{Key Laboratory of Quantum Information, University of Science and Technology of China, Hefei, 230026, China}
\affiliation{CAS Center for Excellence in Quantum Information and Quantum Physics, University of Science and Technology of China, Hefei, 230026, China}

\date{\today}

\begin{abstract}
We consider the phase transition dynamics of a trapped Bose-Einstein condensate subject to Raman-type spin-orbit coupling (SOC). By tuning the coupling strength the condensate is taken through a second order phase transition into an immiscible phase. We observe the domain wall defects produced by a finite speed quench is described by the Kibble-Zurek mechanism (KZM), and quantify a power law behavior for the scaling of domain number and formation time with the quench speed.
\end{abstract}

\maketitle

{\it Introduction}. Non-equilibrium physics is an active area of current research. While fewer tools exist for understanding out-of-equilibrium processes, universal behavior can still emerge. For instance, in the formation of topological defects in symmetry-breaking phase transitions, or in the self-similar growth of domains via defect annealing in phase ordering dynamics.

The Kibble-Zurek mechanism (KZM) \cite{Kibble1976,Zurek1985} is a theory used to describe the formation of defects at a phase transition in terms of the relevant critical exponents. There are many theoretical studies of the KZM in various systems, such as Landau-Zener transitions \cite{Damski2005,Damski2006}, temperature quenching across the Bose-Einstein condensation (BEC) transition \cite{Damski2010}, Ising model \cite{Dziarmaga2005} and various types of spinor condensates \cite{Lee2004,Lamacraft2007a,Saito2007a,Wu2017,Anquez2016,Uhlmann2007}. To date experimental tests of the KZM have been performed in liquid crystals \cite{Chuang1991b}, cold atomic systems \cite{Weiler2008,Clark2016,Labeyrie2016}, linear optical systems \cite{Xu2014} and ion Coulomb crystals \cite{Pyka2013}. A key challenge is to have a system in which there is good control over the rate at which the phase transition is crossed, and where the defects formed are relatively stable and able to be measured.
 
Experiments with cold atoms are able to dynamically engineer interesting single particle properties and control interactions. This makes for a rich system to control and explore phase transition dynamics, particularly since topological defects can be readily detected in experiments \cite{Baumann2011,Chae2012,Parker2013,Campbell2016}. 

In this paper we are motivated by the phenomenal development made in experiments producing spin-orbit coupling (SOC) terms in cold atom systems. We study the phase transition dynamics of a two component BEC with a Raman-type SOC  \cite{lin2011}. As the coupling strength is varied the system undergoes a quantum phase transition where the spin components change from being miscible to immiscible, also accompanied by changes in the momentum distribution. We study this problem in an experimentally realistic case of a quasi-1D harmonic trap using the truncated Wigner method. We find that domain walls defects form separating the spin components in the immiscible phase. The number of these defects follows a scaling law related to the rate that the phase transition is crossed.


{\it Kibble-Zurek Mechanism}. Consider a uniform system crossing a second order phase transition point, and introduce a parameter $\epsilon$ to quantify the distance from the critical point (i.e.~$\epsilon=0$ is the critical point). For instance, $\epsilon$ could be some thermodynamic parameters like temperature (e.g.~\cite{Zurek2009}) or some Hamiltonian control parameter  (e.g.~\cite{Clark2016}). Here we will consider $\epsilon$ to be controlled by the intensity of the Rabi coupling. Near the critical point the correlation length and relaxation time diverge as
\begin{equation}
\xi=\xi_0/|\epsilon|^\nu,\quad
\tau=\tau_0/|\epsilon|^{\nu z},
\end{equation}
respectively, where $\xi_0$ and $\tau_0$ depend on the specific system, while the critical exponents $\nu$ and $z$ are determined by the universality class of the phase transition.

\begin{figure}
\centering
\includegraphics[width=8.5cm,height=6cm]{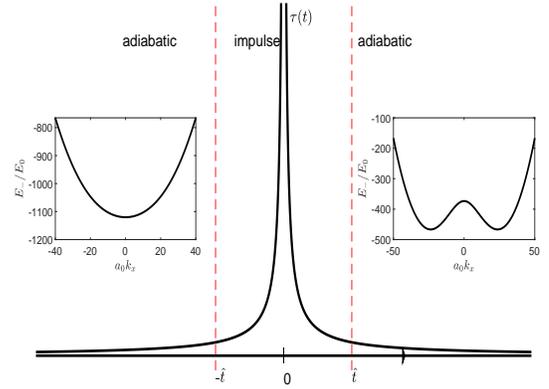}
\caption{Relaxation time from Kibble-Zurek theory. Inset: Lower branch $E_{-}$ of the dispersion of the single-particle part of the SOC Hamiltonian in the absence of trapping for $\Omega=1.5\Omega_c$ (left) and $\Omega=0.75\Omega_c$ (right).  There is one minimum for $\Omega>\Omega_c$ and two degenerate minima for $\Omega<\Omega_c$.}
\label{fig:1}
\end{figure}

According to KZM, when the phase transition is crossed, critical slowing-down intervenes \cite{Zurek2009}: 
as the relaxation time diverges correlations freeze in at a length scale determined by the speed that the system crosses the transition. Thus different parts of the system make different choices for the symmetry broken order parameter and domains of order are produced.

The freezing time $\hat{t}$ is a pivotal quantity in the KZM, defining when the evolution becomes non-adiabatic (see Fig.~\ref{fig:1}). We identify this time by equating the quench time scale $\epsilon/\dot \epsilon$ to the relaxation time
\begin{equation}
\tau(\hat t)=\epsilon(\hat t)/\dot \epsilon(\hat t).
\end{equation}
For a linear quench $\epsilon=t/\tau_q$, where $\tau_q$ is the quench time, we obtain $\hat t=(\tau_0\tau_q^{\nu z})^{\frac{1}{1+\nu z}}$, and the correlation length that freezes in at $\hat{t}$ is
\begin{equation}
\hat \xi=\xi_0({\tau_q}/{\tau_0})^{\frac{\nu}{1+\nu z}},\label{xihat}
\end{equation}
this sets the length scale for defects production as the transition is crossed.

\begin{figure*}[ht]
\centering
\includegraphics[width=18cm,height=8.5cm]{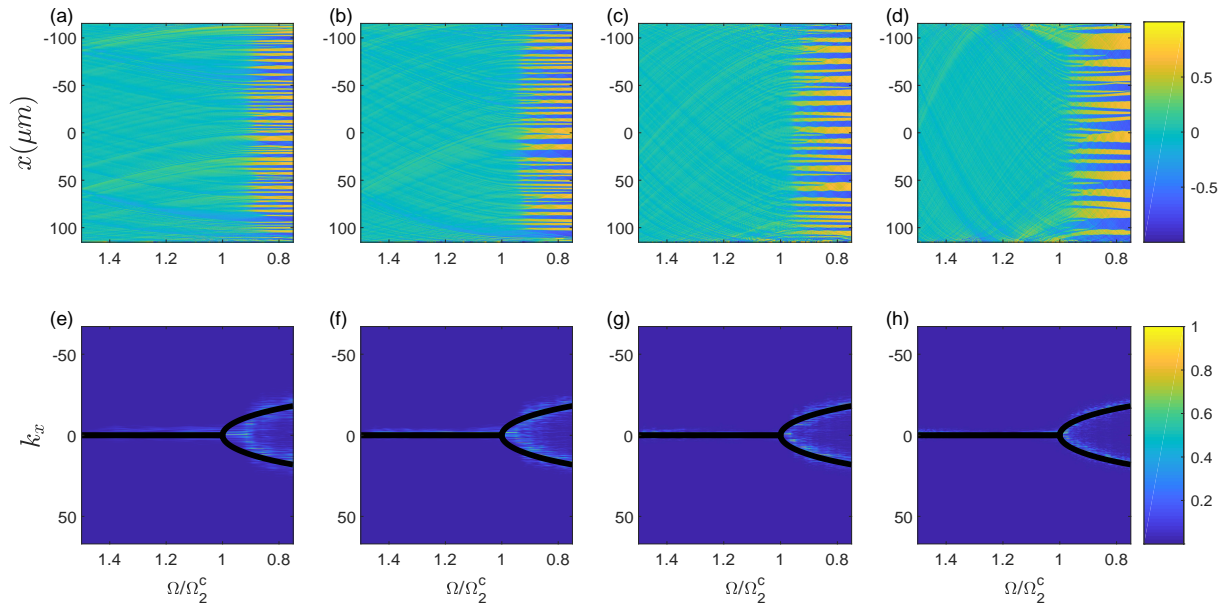}
\caption{Upper panel: local magnetization $j_z(x)$ as function of $\Omega/\Omega_2^c$ for quench time (a) $\tau_q=40$ ms, (b) $\tau_q=80$ ms, (c) $\tau_q=160$ ms and (d) $\tau_q=320$ ms, lower panel: bifurcation of momentum distribution as function of $\Omega/\Omega_2^c$ for quench time (e) $\tau_q=40$ ms, (f) $\tau_q=80$ ms, (g) $\tau_q=160$ ms and (h) $\tau_q=320$ ms. The black solid-curve indicate the instantaneous momentum distribution. We have only plotted the region within the radius $R=R_{TF}$, where $R_{TF}=115 \mu$m is the Thomas-Fermi radius.}
\label{fig:2}
\end{figure*}
{\it Bose-Einstein Condensate with SOC}. The Raman-type SOC has been experimentally realized in cold atom system \cite{lin2011}. Here a pair of lasers couple two internal atomic states denoted by 1 and 2 with Raman coupling strength $\Omega$, and two-photon momentum transfer of  $\hbar k_r$. Here we focus on an elongated quasi-one-dimensional (quasi-1D) system in a harmonic trap of angular frequency $\omega_x$ and with the SOC momentum vector $k_x$ taken to be along $x$.

Choosing $a_0=\sqrt{\hbar/m\omega_x}$, $t_0=1/\omega_x$, and $E_0=\hbar \omega_x$  as the units of length, time and energy, respectively, the dimensionless time-dependent Gross-Pitaevskii equation (GPE) for this system is (e.g~see \cite{lin2011,Ho2011,LiYun2012,Liyun2013}) $i\partial_t\Psi=H\Psi$,
where 
\begin{equation}
H=\begin{pmatrix}
\frac{k_x^2}{2}+\gamma k_x+ I_1 & \tfrac{1}{2}\Omega\\
\tfrac{1}{2}\Omega\ & \frac{k_x^2}{2}-\gamma k_x + I_2
\end{pmatrix}
+\frac{x^2}{2}+\frac{\delta}{2}\sigma_z,\label{eqH}
\end{equation}
 $\Psi=(\psi_1,\psi_2)^T$, $I_j=g_{j1}|\psi_1|^2+g_{j2}|\psi_2|^2$, with $\psi_j$  being  the condensate wave function for spin state $j$ and $k_x$ denoting the quasi-momentum, $\gamma=a_0k_r$ is a dimensionless constant and we have eliminated a constant energy term $\gamma^2/2$ in the Hamiltonian. 

The interactions between the spin components are described by the short ranged  intra-species $\{g_{11},g_{22}\}$ and the inter-species $g_{12}=g_{21}$  coupling constants.  Here $\delta$ is the detuning of the Raman coupling, with $\sigma_z$ the Pauli $z$-matrix. From now on we set $\delta=0$, and take the intra-species interactions to be identical, i.e.~$g_{11}=g_{22}=g$.

The dispersion relations for the single-particle part of the Hamiltonian (\ref{eqH}) in the absence of the harmonic trap are
\begin{equation}
E_{\pm}=\frac{k_x^2}{2}\pm\sqrt{k_x^2\gamma^2+\frac{\Omega^2}{4}}.
\end{equation}
The lower branch $E_{-}$ has one or two minima depending on the value of $\Omega$. As shown in the insets of Fig.~\ref{fig:1}, for $\Omega>\Omega_c\equiv 2\gamma^2$, the minimum is at $k_x=0$, and for $\Omega<\Omega_c$, the minima are at $k_x=\pm k_0$, where $k_0=\sqrt{\gamma^2-\frac{\Omega^2}{4\gamma^2}}$.

Including interaction effects this system exhibits three different phases that are accessible under appropriate conditions. Denoting the total density as $n=|\psi_1|^2+|\psi_2|^2$,  and defining $G_\pm=\frac{1}{4}n({g\pm g_{12}})$, the system can access all three phases if the density is less than the critical value $n^{c}= {\gamma^2G_+}/{2gG_-}$:

(1) {\it Stripe phase}: The condensate atoms are in an  superposition of $\pm k_0$ momentum states, occurring for $\Omega<\Omega_1^{c}$. In this state the overall magnetization of the state is zero, i.e.~$M_z\equiv \int dx(|\psi_1|^2-|\psi_2|^2)=0$.

(2)  {\it Plane-Wave phase}:  The condensate atoms prefer to occupy either of the $\pm k_0$ momentum states when $\Omega_1^{c}<\Omega<\Omega_2^{c}$. This phase breaks the spin symmetry with the magnetization $M_z$ being non-zero, taking the value $M_z=\pm k_0N$, where $N=\int dx\,n$.

(3) {\it Zero-Momentum Phase}: When $\Omega>\Omega_2^{c}$, $k_0=0$ and the condensate atoms occupy the zero momentum mode. This state has $M_z=0$, although exhibits transverse magnetization.

In describing the phases above we have introduced $\Omega_1^{c}\equiv\sqrt{\frac{8G_-(\gamma^2+G_+)(\gamma^2-2G_-)}{G_++2G_-}}$ and $\Omega_2^{c}\equiv 2(\gamma^2-2G_-)$. We also note that the transition from zero-momentum phase to plane-wave phase is a second order phase transition \cite{Liyun2013,Hamner2014} and we can explore the KZM in this scenario. 

The system breaks a $Z_2$ symmetry and chooses between the two possible ground states when it crosses from zero-momentum phase to plane-wave phase. These ground states can be distinguished by their momenta and spin composition (magnetization). When phase transition is crossed at finite rate, then the ground state choice is made locally in the system, giving rise to domains. We characterize the local order using normalized magnetization density 
\begin{align}
j_z(x)=\frac{|\psi_1(x)|^2-|\psi_2(x)|^2}{|\psi_1(x)|^2+|\psi_2(x)|^2},
\end{align}
which will gain a non-zero value of either $\pm k_0/\gamma$ in the plane-wave phase.

{\it Quench dynamics of Trapped BEC with SOC}. As described above, $G_{-}=n(x)(g-g_{12})/4$ depends on the position if we consider a system with trap potential, so the critical $\Omega_2^c(x)=2[\gamma^2-2G_{-}(x)]$ depends on position, where $n(x)$ is taken to be the initial total density. To implement a  quench we ramp down $\Omega$ from an initial value of $\Omega_i=1.5\Omega_2^c(x)$ in the zero-momentum phase to $\Omega_f=0.75\Omega_2^c(x)$ in the plane-wave phase. In the ramp $\Omega$ changes linearly in time over a time interval $\tau_q$ that we take to define the quench time, i.e.~$\Omega(t)=\max\left([\Omega_f,\Omega_i-(\Omega_i-\Omega_f) t/\tau_q)]\right)$, where we choose $t\geq 0$. Here we study quench times ranging from 10 ms to 1000 ms.

We consider a system of $N=10^4$ $^{87}$Rb atoms in a trap with frequency $\omega_x=2\pi\times 5$ Hz, $\omega_y=\omega_z=2\pi\times 2$ kHz, and  interspecies interaction strength of $g_{12}=1.05g$, with $g=4\pi a_s\hbar^2/m$, where $a=100.86\,a_0$ and $a_0$ is the Bohr radius. We consider the SOC to be produced by $\lambda=784$ nm  lasers crossed at an angle of $\pi/2$, so that the recoil momentum is $k_r=\sqrt{2}\pi/\lambda$. 
 Under this condition, $G_-=n(g-g_{12})/2= ng_s/2$ (where $g_s=g-g_{12}$) is small and negative, this gives $\Omega_1^c<0$, so we can exclude the stripe phase from our analysis.  This choice also means that the critical point identified in the uniform system $\Omega_2^c$  is weakly dependent on density, and thus inhomogeneous density of the trapped system should not strongly affect the results, which is very different from the system that was considered in \cite{Sabbatini2011}. Based on the parameters we considered, the critical Rabi coupling at the center is about $99.6\%$ of that at the edge.

To simulate the quench dynamics we use the truncated-Wigner method \cite{Blakie2008}, whereby initial noise is added to the Bogoliubov quasi-particle modes to simulate the effects of vacuum fluctuations. The initial condensate (at $\Omega=1.5\Omega_2^c$) is obtained by imaginary time propagation, and then the quasi-particle modes are calculated by numerical diagonalization and used to add noise to construct the initial field (see the Appendix for more details). The simulation is then performed by evolving the initial field in real time with the GPE as $\Omega$ is linearly ramped to effect the quench.

For each quench time $\tau_q$ we conduct 100 trajectories of the truncated Wigner simulations which we use to compute statistics. In each trajectory a different sampling of noise is used to construct the initial condition, and has the effect of providing a different seed for the growth of the symmetry breaking domains during the quench. In the upper panel of Fig.~\ref{fig:2} we show single realization of $j_z(x)$ for several quench times. We can clearly see the typical domain sizes and the number of domains vary with time. The domains are less prominent towards the edge of the condensate because the density is lower there. Noise or thermal excitations from the quench can be more important in the low density wings making the identification of the domains difficult in this region, so we only count the domains number within the Thomas-Fermi radius. As described before, the single-particle dispersion transitions from having a single minimum at $k_0=0$ to having two degenerate minima at $k=\pm k_0$ in the quench process. According to KZM, there will be a delay for the momentum bifurcation for finite quench speed, and this delay effect can be seen in the lower panel of Fig.~\ref{fig:2} and we can use this effect to determine the domain formation time as we will discuss below.

In order to extract the power law, we count the number of domains $N_q$ by determining how many times $j_z(x)$ crosses zero within some region. To make sure the growth of domains ceased to increase, we take $N_q$ to be the mean zero-crossing number from time $t=\tau_q$ to $t=\tau_q+20$ ms.

\begin{figure}
\centering
\includegraphics[width=9cm,height=6cm]{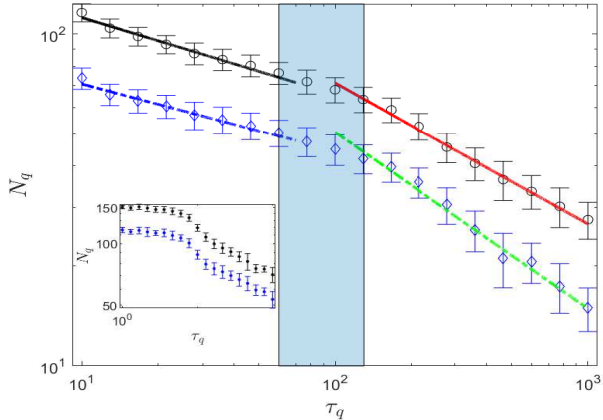}
\caption{Domains number $N_q$ as function of quench time $\tau_q$ (in units of ms). Circles and diamonds with error bar correspond to numerical results for region $R=R_{TF}$ and $R=0.7R_{TF}$, respectively. Black-solid line and Blue-dash-dotted line are the fittings for $\tau_q$ between 10 ms and 50 ms. Red-solid line and green-dash-dotted line are the fittings for $\tau_q$ between 150 ms and 1000 ms. A power law $N_q=\tau_q^{-\alpha}$ fits for the data points of the black-solid line and the blue-dash-dotted line gives $\alpha=0.23\pm 0.04$ for $R=R_{TF}$, and $\alpha=0.20\pm 0.04$ for $R=0.7R_{TF}$. The same law for the data points of the red-solid line and the green-dash-dotted line gives $\alpha=0.43\pm 0.03$ for $R=R_{TF}$, and $\alpha=0.53\pm 0.06$ for $R=0.7R_{TF}$. Inset: domains number $N_q$ for quench time 1 ms $<\tau_q<$ 100 ms. The shaded area is the crossover region.}
\label{fig:3}
\end{figure}

We can explore the phase transition scaling by seeing how the number of domains produced in the quench depends on the quench time. The number of domains will scale as $N_q\sim 2R/\hat{\xi}\sim\tau_q^{-\nu/(1+\nu z)}$, where $R$ is the region's radius  and $\hat{\xi}$ is correlation length at frozen time $\hat{t}$ (\ref{xihat}), which depends on the quench time. A power-law fit to the results summarized in Fig.~\ref{fig:3} allows us to extract the exponent $\frac{\nu}{1+\nu z}=0.23\pm 0.04$ for $R=R_{TF}$ and $\frac{\nu}{1+\nu z}=0.20\pm 0.04$ for $R=0.7R_{TF}$. Also, the results appear to begin deviating from power-law behavior when quench time $\tau_q$ is larger than 150 ms or the radius of region $R<R_{TF}$ , the scaling exponent under these conditions is not consistent with the prediction of KZM,  we conjecture this is because the system has reached the adiabatic region and because of the trap potential we are using. We denote the shaded-area in Fig.~\ref{fig:3} as crossover region which connects the KZM region and the adiabatic region.

\begin{figure}
\centering
\includegraphics[width=9cm,height=6cm]{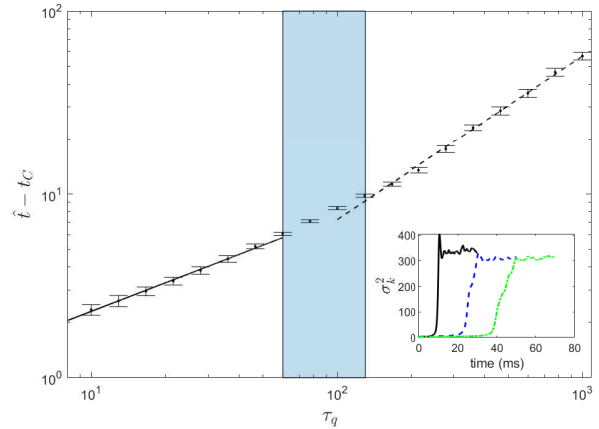}
\caption{$\hat{t}-t_C$ as function of quench time $\tau_q$ (in units of ms), this data is extracted from the second moment of momentum distribution. Black-solid line is the fitting curve $\ln(\hat{t}-t_C)\sim -\frac{\nu z}{1+\nu z}\ln \tau_q$, the fitting parameter is $\frac{\nu z}{1+\nu z}=0.52\pm 0.03$ for quench time $\tau_q$ between 10 ms and 50 ms, and $\frac{\nu z}{1+\nu z}=0.89\pm 0.05$ for quench time $\tau_q$ between 150 ms and 1000 ms. Inset: Second moment of momentum distribution $\sigma_k^2$ as function of time for quench time $\tau_q=10$ ms (black-solid line), $\tau_q=30$ ms (blue-dashed line), and $\tau_q=50$ ms (green-dash-dotted line). The shaded area is the crossover region.}
\label{fig:4}
\end{figure}

After the system crossed the phase transition point with finite speed, according to KZM, there will be two peaks in the momentum distribution, we calculated the time-dependent second moment of the momentum distribution 
$\sigma_k^2=\int k^2 n(k){\rm d}k/\int n(k){\rm d}k-(\int k n(k){\rm d}k/\int n(k){\rm d}k)^2$, where $n(k)=|\phi_1(k)|^2+|\phi_2(k)|^2$ and $\phi_j(k)$ are the momentum distribution of wavefunction $\psi_j(x)$. In the inset of Fig.~\ref{fig:4}, we can clearly see a sharp slope, and we choose a threshold 50 to indicate the domains formation time $\hat{t}$.

In Fig.~\ref{fig:4} we plot $\hat{t}-t_C$ as a function of the quench time $\tau_q$ (see the Appendix) and we extract another exponent $\frac{\nu z}{1+\nu z}=0.52\pm 0.03$. From the above two scaling laws, we can extract the scaling exponents $\nu=0.48$ and $z=2.26$.

{\it KZM and Adiabatically Area}. For quench time $\tau_q<10$ ms, the domains number saturated to a constant value as we can see from the inset of Fig.~\ref{fig:3}, this constant value is reaching the maximum domains number allowed by the system. In this region, the KZM will not give the correct value due to the miscounting of domain number. We also find that, when the quench time $\tau_q> 150$ ms, the domains' number and frozen time will not follow the prediction of KZM, as seen in Fig.~\ref{fig:3} and Fig.~\ref{fig:4}. We conjecture that the system is nearly adiabatic when quench time $\tau_q>150$ ms, and our findings suggest that the KZM only works for weak trap potential and fast enough quench times \cite{Saito2013,Campo2011},.


{\it Discussion and Conclusion}. Stimulated by recent experimental and theoretical work concerning on KZM, we studied KZM in trapped BEC with SOC within the framework of truncated-Winger GPE. We have observed domains formation in the quench process, we also extracted two power laws from the formation time and domains number data. We get two scaling exponents $\nu$ and $z$. In our scheme, the defects are formed in spatial space which makes it is easier to detect in experiment. Compare to the study of KZM in Landau-Zener system \cite{Damski2005} or in Ising model \cite{Dziarmaga2005}, our system includes the interaction which makes it a many-body system instead of single-particle system. Also in order to make the system more experimentally realistic we have includes a harmonic trap potential. There still lies an open question: is the crossover region which connects the KZM region to the adiabatic region universal and how can we identify this region analytically and experimentally? 

In the experiment of Ref. \cite{Erhard2004,Tojo2010}, they used two hyperfine states $|1\rangle=|F=1,m_F=+1\rangle$ and $|2\rangle=|F=1,m_F=-1\rangle$ of $^{87}$Rb as the two pseudo-spin states. These two hyperfine states have a property $g_{11}\approx g_{22}$ and they can tune $g_{12}$ to be very close to $g_{11}$. For $N=10^4$, $\omega_x=2\pi \times 5$ Hz, $\omega_y=\omega_z=2\pi\times 2$ kHz \cite{Sabbatini2011}, the Thomas-Fermi radius is  $R_{TF}\approx 115$ $\mu$m, healing length at the trap center is about $\xi=\sqrt{\frac{\hbar^2}{2mn(0)g}}\approx 0.2$ $\mu$m, the spin healing length $\xi_s=\sqrt{\frac{\hbar^2}{2mn(0)|g_s|}}$ is about $4.5\xi$, there are about $N_d^{max}\sim 250$ defects maximum in our system. So this scheme is feasible to be realized in experiment with current technology.

{\it Note added}. Another paper was posted online \cite{Ye2018} when we were conducting our project. Quenched dynamics of Raman-type SOC in uniform BEC was considered there. Here we considered a more practical model of trapped BEC.


The authors acknowledge very valuable help by Blair Blakie on numerical methods and the manuscript, we also thank Xiangfa Zhou and Chuanwei Zhang for valuable discussions. This work is supported by National Key R\&D Program (No. 2016YFA0301300 and No. 2016YFA0501700), National Natural Science Foundation of China (No. 11674306 and No. 61590932), and Anhui Initiative in Quantum Information Technologies.

%

\section{Appendix}
\subsection{Bogoliubov-de Gennes Equation}
In the following we use Bogoliubov-de Gennes (BdG) method \cite{BdG} to get the excitation energies and the corresponding wave functions.

For the ground state wave function $\Phi_g$ of SOC BEC, we have the stationary GPE
\begin{equation}
\mu \Phi_g=H\Phi_g,
\end{equation}
where $\Phi_g=(\phi_{1g},\phi_{2g})^{T}$. We assume the time-dependent wave function is
\begin{equation}
\Psi=[\Phi_g+\delta\Psi]\exp{(-i\mu t)},
\end{equation}
where $\delta\Psi$ is the fluctuation.

Substitute above wave function into time-dependent Gross-Pitaevskii equation and make use of the stationary GPE, we get two equations on $\delta\Psi$

\begin{widetext}

\begin{equation}\label{eq:bdg1}
i \frac{\partial}{\partial t}\delta\psi_1=
[\frac{k_x^2}{2}+\gamma k_x+V_{trap}+g_{12}|\phi_{2g}|^2+2g|\phi_{1g}|^2-\mu]\delta\psi_1+g\phi_{1g}^2\delta\psi_1^{*}+[g_{12}\phi_{1g}\phi_{2g}^{*}+\frac{\Omega}{2}]\delta\psi_2+g_{12}\phi_{1g}\phi_{2g}\delta\psi_2^{*}
\end{equation}
and
\begin{equation}\label{eq:bdg2}
i \frac{\partial}{\partial t}\delta\psi_2=
[\frac{k_x^2}{2}-\gamma k_x+V_{trap}+g_{12}|\phi_{1g}|^2+2g|\phi_{2g}|^2-\mu]\delta\psi_2+g\phi_{2g}^2\delta\psi_2^{*}+[g_{12}\phi_{1g}^{*}\phi_{2g}+\frac{\Omega}{2}]\delta\psi_1+g_{12}\phi_{1g}\phi_{2g}\delta\psi_1^{*}.
\end{equation}
\end{widetext}

We choose the following excitation form \cite{BEC_Superfluid}
\begin{equation}\label{eq:exci}
\left\{
\begin{aligned}
\delta\psi_1 & = & u_1(x)\exp{(-i\omega t)}-v_1^{*}\exp{(i\omega t)} \\
\delta\psi_2 & = & u_2(x)\exp{(-i\omega t)}-v_2^{*}\exp{(i\omega t)}.
\end{aligned}
\right.
\end{equation}
Substitute Eq. (\ref{eq:exci}) into Eq. (\ref{eq:bdg1}) and Eq. (\ref{eq:bdg2}) and compare the coefficients, we get the BdG matrix equation for $(u_1(x),v_1(x),u_2(x),v_2(x))^{T}$

\begin{equation}\label{eq:matrix}
M_{BdG}
\begin{pmatrix}
u_1\\
v_1\\
u_2\\
v_2
\end{pmatrix}
=\omega\begin{pmatrix}
u_1\\
v_1\\
u_2\\
v_2
\end{pmatrix}
\end{equation}
in which $M_{BdG}$ is the BdG matrix
\begin{equation}
\begin{pmatrix}
A+B & -D & F & -E \\
D^* & -A^*-B^* & E^* & -F^* \\
F^* & -E & A+C & -G \\
E^* & -F & G^* & -A^*-C^*
\end{pmatrix}.
\end{equation}
where $A=\frac{k_x^2}{2}+\gamma k_x$, $B=V_{trap}+2g|\phi_{1g}|^2+g_{12}|\phi_{2g}|^2-\mu$, $C=V_{trap}+2g|\phi_{2g}|^2+g_{12}|\phi_{1g}|^2-\mu$, $D=g\phi_{1g}^2$, $G=g\phi_{2g}^2$, $E=g_{12}\phi_{1g}\phi_{2g}$, $F=\frac{\Omega}{2}+g_{12}\phi_{1g}\phi_{2g}^*$.

Substitute the ground wave function $\{\phi_{1g}, \phi_{2g}\}$ into Eq. (\ref{eq:matrix}), then diagonalize the BdG matrix to get the collective excitation energies $\{\omega_j\}$ and the corresponding excitation wave functions $\{u_{1j},v_{1j},u_{2j},v_{2j}\}$.

After we get the excitation energies and excitation wave functions, we form the initial wave function for the real time evolution as following
\begin{equation}\label{eq:initial_w}
\begin{aligned}
\psi_1(x)= \psi_{1g} + \sum_{j=1}[\beta_ju_{1j}(x)e^{-i\omega t}-\beta_j^*v_{1j}^*(x)e^{i\omega t}]\\
\psi_2(x)= \psi_{2g} + \sum_{j=1}[\alpha_j u_{2j}(x)e^{-i\omega t}-\alpha_j^*v_{2j}^*(x)e^{i\omega t}]
\end{aligned}
\end{equation}
the coefficients $\{\alpha_j,\beta_j\}$ are random numbers sampled from the Wigner distribution for zero-temperature thermal state \cite{QuantumNoise}, i.e.
\begin{equation}\label{eg:wigner_dis}
W(\alpha,\alpha^*)=\frac{2}{\pi}\exp{(-2|\alpha|^2)}.
\end{equation}

\subsection{Quench Parameters}
In our simulation, the quench parameter is defined as
\begin{equation}
\epsilon=\frac{\Omega-\Omega_c}{\Omega_c},
\end{equation}
where $\Omega_c=2\gamma^2$ and $\Omega$ is linearly quenched from $\Omega_i=1.5\Omega_c$ to $\Omega_f=0.75\Omega_c$, so we have
\begin{equation}
\Omega(t)=\frac{\Omega_f-\Omega_i}{\tau_q}t+\Omega_i,
\end{equation}
from above we can get the time $t_0$ that $\Omega(t_0)=\Omega_c$
\begin{equation}
t_0=\frac{\Omega_c-\Omega_i}{\Omega_f-\Omega_i}\tau_q.
\end{equation}

From the main text, we know that the correlation length and relaxation time satisfy $\tau=\tau_0/|\epsilon|^{\nu z}$ and $\xi=\xi_0/|\epsilon|^{\nu}$. We choose $t=t_0+t^{\prime}$, then
\begin{equation}
\epsilon(t)=\epsilon(t_0+t^{\prime})\equiv \eta(t^{\prime}),
\end{equation}
this gives us
\begin{equation}
\tau=\tau_0/|\eta|^{\nu z}.
\end{equation}

The frozen time is defined as $\tau(t^{\prime})=\frac{\eta(\hat{t}^{\prime})}{\dot\eta(\hat{t}^{\prime})}$, we get
\begin{equation}
\hat{t}^{\prime}=(\frac{\Omega_c}{\Omega_i-\Omega_f})^{\nu z/1+\nu z}\tau_0^{\frac{1}{1+\nu z}}\tau_q^{\frac{\nu z}{1+\nu z}},
\end{equation}
Finally, we have
\begin{equation}
\hat{t}-t_0=(\frac{\Omega_c}{\Omega_i-\Omega_f})^{\nu z/1+\nu z}\tau_0^{\frac{1}{1+\nu z}}\tau_q^{\frac{\nu z}{1+\nu z}}\propto \tau_q^{\frac{\nu z}{1+\nu z}},
\end{equation}
and
\begin{equation}
|\Omega(\hat{t})-\Omega_c|=\Omega_c(\frac{\Omega_c}{\Omega_i-\Omega_f})^{\frac{-1}{1+\nu z}}\tau_0^{\frac{1}{1+\nu z}}\tau_q^{-\frac{1}{1+\nu z}}\propto \tau_q^{-\frac{1}{1+\nu z}}.
\end{equation}

\end{document}